# Optical Vortices during a Super-Resolution Process in a Metamaterial

G. D'Aguanno[*(1)], N. Mattiucci[(1)], M. Bloemer[(1)], and A. Desyatnikov[(2)]

*(1) Dept. of the Army, Charles M. Bowden Facility, Bldg. 7804
Research, Development, and Engineering Command, Redstone Arsenal, AL 35898,USA*

*(2) Nonliner Physics Centre and Laser Physics Centre, Centre for Ultrahigh-bandwidth Devices for Optical Systems (CUDOS), Research School of Physical Sciences and Engineering, Australian National University, Canberra, Australian Capital Territory 0200, Australia*

**Abstract**

We show that a super-resolution process with 100% visibility is characterized by the formation of a point of phase singularity in free space outside the lens in the form of a saddle with topological charge equal to -1. The saddle point is connected to two vortices at the end boundary of the lens, and the two vortices are in turn connected to another saddle point inside the lens. The structure saddle-vortices-saddle is topologically stable. The formation of the saddle point in free space explains also the negative flux of energy present in a certain region of space outside the lens. The circulation strength of the power flow can be controlled by varying the position of the object plane with respect to the lens.

---

[*] Corresponding author: giuseppe.daguanno@us.army.mil or giuseppe.daguanno@gmail.com
Fax: 001-256-8422507; Tel.: 001-256-8429815

Vortices are ubiquitous in nature from the macroscopic to the microscopic world. Tornados and hurricanes [1] or whirlpools [2] are maybe the most common examples at macroscopic distances. At microscopic distances, vortices observed in a superfluid HeII [3] and in a Bose-Einstein condensate of alkali [87]Rb atoms [4-5]. In optics vortices have been observed in the near field diffracted by an array of subwavelength apertures [6-9], in lasers [10], in optical fibers or in systems that create caustics or speckle [11], in computer generated holograms [12] or spiral phase plates [13], during the propagation of the light in self-defocusing, cubic media [14-15] or in quadratic materials [16]. Their applications include: free space interconnection of electronic components [17], optical trapping of viruses and bacteria [18] and optical trapping of particles [19], quantum information and quantum cryptography [20-21], fluorescence microscopy with nanoscale resolution [22], extra-solar planet detection [23-24]. Optical vortices are based on the appearance of phase singularities (also called: "phase dislocations") whenever the field intensity vanishes, as pointed out in the seminal paper by Nye and Berry[25]. Let us consider for simplicity the expression of the time-averaged Poynting vector for a monochromatic field in a two-dimensional geometry: $\vec{S}(x,z) = (1/2)\operatorname{Re}[\vec{E}(x,z) \times \vec{H}^*(x,z)]$ where $\vec{E}(x,z)$ and $\vec{H}(x,z)$ are the complex amplitudes of the electric and magnetic fields respectively. The phase of the Poynting vector can be extracted from its components through the following equations: $\sin\Phi_S(x,z) = S_z(x,z)/|\vec{S}(x,z)|$ and $\cos\Phi_S(x,z) = S_x(x,z)/|\vec{S}(x,z)|$. Now, if we assume that at some point in space $|\vec{S}(x,z)| = 0$ (black spot), then the phase of the Poynting vector will be, of course, not defined (singular) at that point and in the neighborhood of that point there can be three possible situations: a) a left-handed or right-handed power flow (optical vortex), b) a power flow in the form of a source or sink, c) a power flow in the form of a saddle [25-26].

Obviously in free-space a power flow in the form of a source or a sink is excluded. Due to the appearance of those phase singularities, the branch of optics that studies vortices is sometime referred to as "singular optics" and singular optics is emerging as a new and exciting chapter of modern optics [27]. A quantity that is useful in the study of optical vortices is the so-called "topological charge" which can be defined for the Poynting vector as [28]: $s = (1/2\pi)\oint_C \nabla \Phi_S \cdot d\vec{r}$ where the path integral is taken counterclockwise along any closed line $C$ surrounding the point of phase singularity. The topological charge is in general an integer number because the phase varies of multiple of $2\pi$ around the singular point and it counts basically the number (with its algebraic sign) of phase jumps along the closed line $C$ associated with the helical structure of $\Phi_S$.

In the past few years, negative index materials (NIMs), i.e. materials that have simultaneously negative permittivity ($\varepsilon$) and permeability ($\mu$), have been the subject of intense theoretical and experimental investigations [29-30]. One of the most important applications is the possibility to use them to construct a "perfect" lens, i.e. a lens that can also focus the evanescent near-field components of an object, as pointed out by Pendry several years ago in his seminal paper [29]. In 2005 the first NIMs operating in the visible regime were reported [31-32] and shortly after a silver-based NIM operating at telecommunication wavelengths was theoretically studied [33] and experimentally realized [34]. One serious issue that plays a detrimental role toward the achievement of a super-resolving lens is the fact that in currently available meta-materials the absorption or scattering losses are still very high. A much simpler super-resolving lens can be obtained by using one-dimensional metallo-dielectric multilayer structures [35-38] in which low group velocity surface plasmon modes are excited for TM polarization of the light. Those metallo-dielectric lenses have only the permittivity (ε) negative due to the presence of the metal layers and therefore they mimic a NIM only for TM polarization of the light [39, 35-38], nevertheless they retain many salient characteristics of a true NIM as regards super-resolution purposes, and, more important, they have the advantage of low losses in the visible range. In a recent publication [39] we have shown that broadband super-resolution can be achieved in a metallo-dielectric lens made of 5.5 periods of Ag/GaP (22 nm/35 nm) with 17 nm thick GaP anti-reflection coatings on the entrance and exit faces for a total length L=341nm. The lens maintains a good transparency (~60% of the input power is transmitted by the lens) over the super-resolving range 500nm-650nm. For an incident wavelength of 532nm the lens is able to resolve two slits (40nm wide with center to center distance of 140nm) with 100% visibility at 50nm beyond the end face of the lens. The slits are placed at the entrance of the lens (but in free space). In Fig.1(a) we show $S_z(x,z)$ during the super-resolution process and in Fig.1(b) we show $S_z(x,z\cong391nm)$ where the visibility approaches 100%. The black spot indicates the point of coordinate (x=0,z≅391nm) where $S_z$=0. At points in the vicinity of x=0, z≅391nm, $S_z\neq0$. Fig.2(a) shows that not only $S_z$=0 but also $S_x$=0 at the point (x=0,z≅391nm) and therefore we can expect the presence of a point of phase singularity around this point (black spot). In Fig.2(b) we show the phase $\Phi_S$ around the black spot. In this case we have just one phase jump of 2π along a closed line *C* surrounding the black spot and therefore the topological charge is *s=-1* when *C* is oriented counterclockwise. In Fig.3 we show the vector field $\vec{S}$ around the black spot. The black spot is a saddle point and the power flow is characterized by four perfectly symmetric regions of circulation. The energy is flowing toward the black spot along the x-direction, while it is flowing away from the black spot along the z-direction. Note that the flux of energy is negative in the region of free space delimited by the following boundaries: ($z_{Min} \cong 341nm$, $z_{Max} \cong 391nm$, $x_{Min} \cong -18nm$, $x_{Max} \cong 18nm$). A negative flux of energy in free space may appear counterintuitive but is instead perfectly explained by the formation of the saddle point at (x=0, $z \cong 391nm$). This negative flux of energy has been also reported in Ref.[40], although there it had been ascribed to the finite transverse dimension of the lens, while it is clear now that it is due to the formation of a saddle point. The strength of the circulation of the power flow in air and the position of the saddle point can be controlled by varying *a*, i.e. the distance of the object plane with respect to the input face of the lens. In particular the position of the black spot becomes closer to the end of the lens and at the same time the strength of the circulation of the power flow decreases when the distance between the object plane and the beginning of the lens increases. We have also calculated the diffraction from the two slits in free space (i.e. without the metallo-dielectric lens in place). In this case there is no saddle point on the z-axis at x=0 and no negative energy flux. Therefore the onset of the saddle point must be ascribed to the presence of the super-lens and to the fact that 100% visibility is achieved. In Fig.4 we show the circulation of the Poynting vector $\left(\nabla \times \vec{S}\right)_y$ [41] in free space at $z \cong 343nm$ (i.e. in free space immediately after the end of the lens) as function of *a,* by varying *a* in step of 10nm. As we will see in a moment, the peaks and the valleys of the circulation are related respectively to the presence of left-handed (LV) and right handed (RV) vortices at the boundary of the lens.

Usually the saddle must be connected to other dislocations, and in fact in Fig.5 we show that the saddle point (S1) in our case is connected to two vortices, left-handed (LV1) and right handed (RV1) respectively, which are centered at the last Ag/GaP interface in a symmetric position with respect to the saddle point. The two vortices are in turn connected to another saddle point (S2) inside the lens. The structure S1-LV1-RV1-S2 is a manifestation of the super-resolution process with 100% visibility. In order to address the issue of the stability of the structure, we calculate the Poynting vector outside the lens in the case of two slightly asymmetric slits. The first slit is now positioned between $x1_-$=-100nm and $x1_+$=-50nm (therefore the width of the slit is now 50nm instead of 40nm) while the position and the width of second slit remain unchanged. The calculations show that the structure S1-LV1-RV1-S2 remains practically unchanged with just a slight modification in its location proving therefore the topological stability. We would like to underline that in this Letter we have focused our attention only on the structure S1-LV1-RV1-S2 which characterizes the super-resolution with 100% visibility. As a matter of fact, the structure of the phase dislocations inside the super-lens is much more complicated and far richer, although a detailed mapping of all the phase dislocations is outside the scope of the present Letter. The calculations presented here have been performed using the angular spectrum representation technique [42] in conjunction with a matrix transfer technique [43]. The method, given its intrinsic analytical nature, avoids any numerical problem.

In conclusion we have demonstrated that a super-resolution process with a 100% visibility is accompanied by the formation of a structure S1-LV1-RV1-S2 which is topologically stable. We expect that our findings may have important applications in the field of optical interconnections at the nanoscale level. We hope also that our results may stimulate the finding of further connections between the fascinating fields of "singular optics" and "metamaterials". G.D. and N.M. thank the NRC for financial support. G.D. and A.D. thank Yuri Kivshar for helpful discussions.

**Figure Captions**

**Fig.1:** (a) $S_z$ (arbitrary units) in the x-z plane. The black dashed-line indicates the position of the end of the lens, the red line indicates the position of the image plane. (b) Section of $S_z$ (red curve) at the image plane (z≅391nm) where 100% visibility is achieved. Superimposed the position of the two slits (black dashed-line)

**Fig.2:** (a) 3-D plot of $|\vec{S}| = \sqrt{S_x^2 + S_z^2}$ in the x-z plane in a small region around the black spot. (b) Structure of the phase of the Poynting vector ($\Phi_S$) around the black spot.

**Fig.3:** Arrow representation (red) of the vector field $\vec{S}$ in a small region around the saddle point (black spot). The black arrows are just for illustrative purposes. For a better view the magnitude of the Poynting vector has been magnified 10 times with respect to the values reported in Fig.1 and in Fig.2(a).

**Fig.4:** $(\nabla \times \vec{S})_y$ (arbitrary units) vs. x (microns) at z≅343nm for different distances *a* of the object plane from the input surface of the lens.

**Fig.5:** Arrow representation of the unit-vector: $\vec{S}/|\vec{S}|$. The saddle point (S1) is connected to two vortices, right-handed (RV1) and left-handed (LV1) respectively, centered at the last Ag/GaP interface. These vortices are in turn connected to a second saddle point (S2) inside the lens centered at the penultimate GaP/Ag interface. The black arrows are just for illustrative purposes. They sketch the flux of the energy. Note in particular the negative flux of energy directed toward the end of the lens.

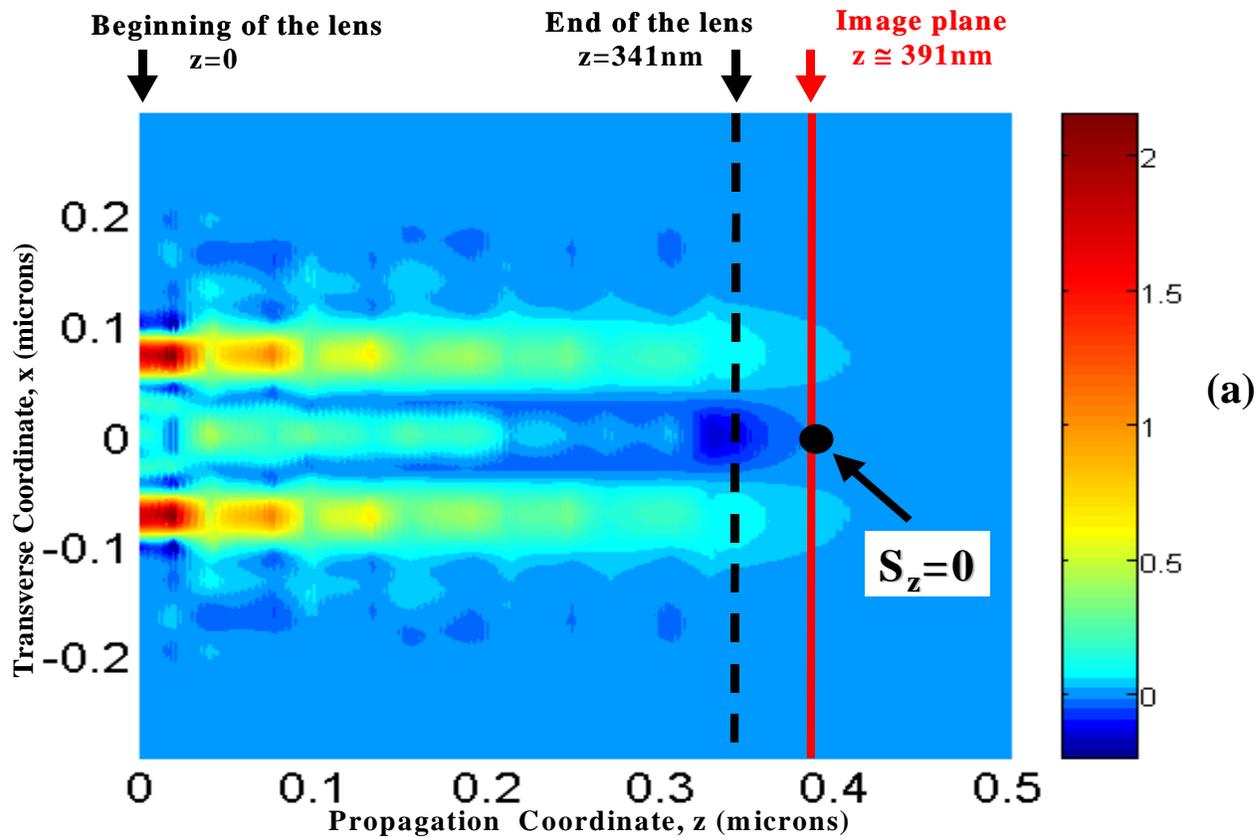

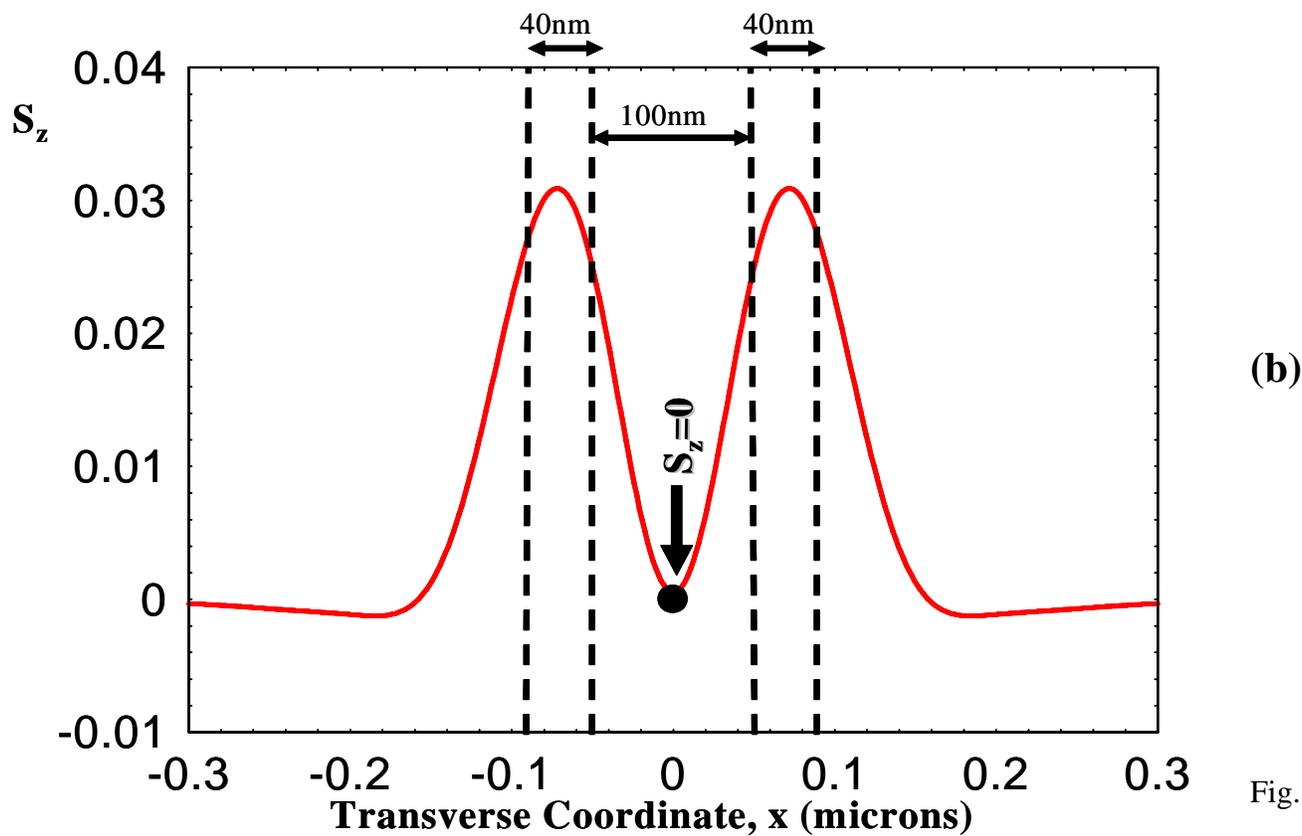

Fig.1

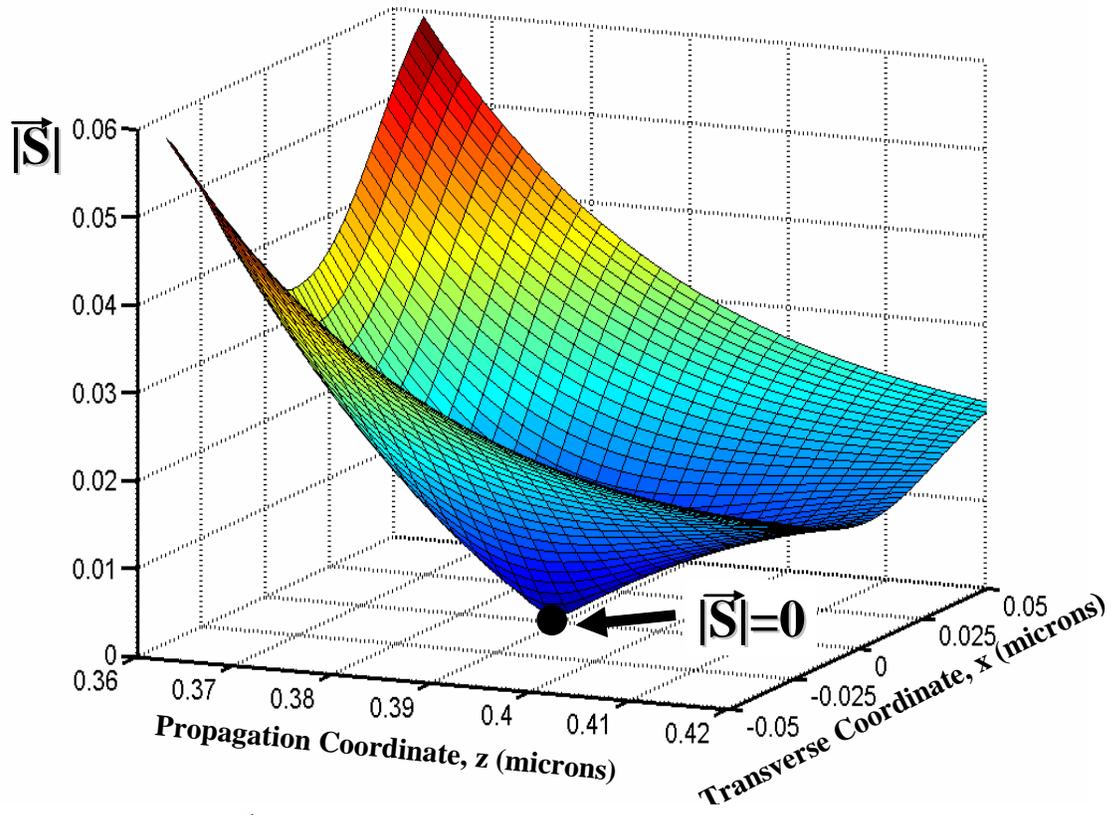

**Topological charge:** $\dfrac{1}{2\pi}\oint_C \nabla \Phi_s \cdot d\vec{r} = -1$

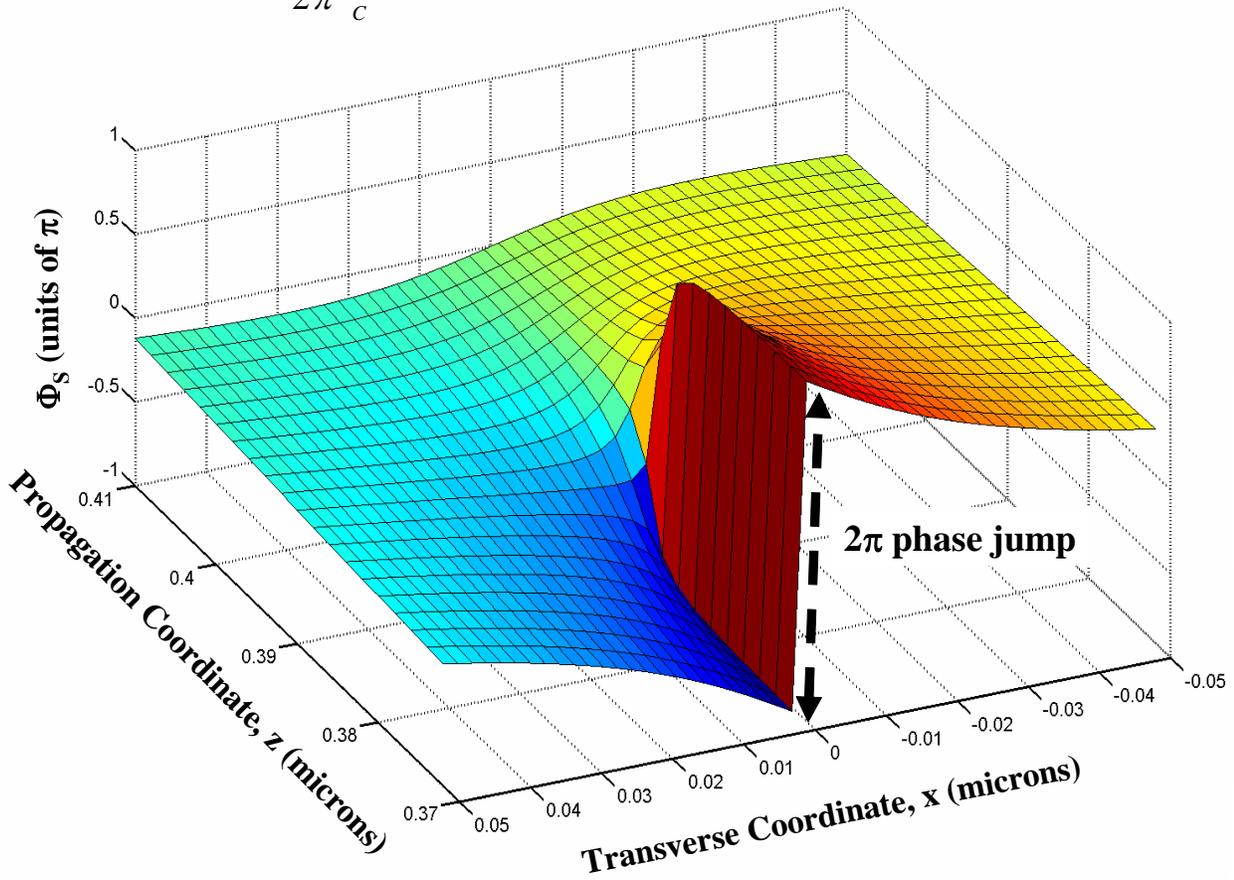

Fig.2

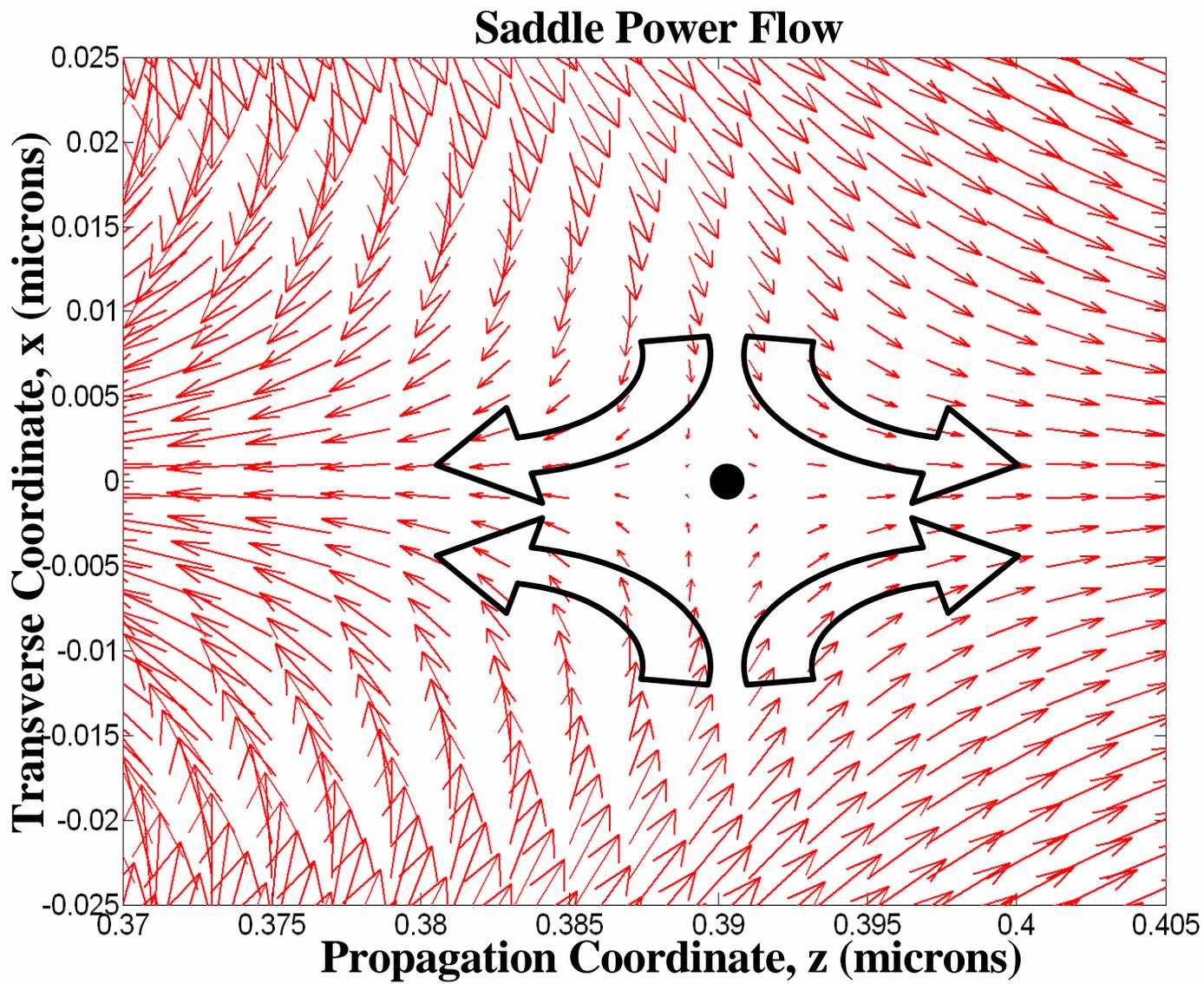

Fig.3

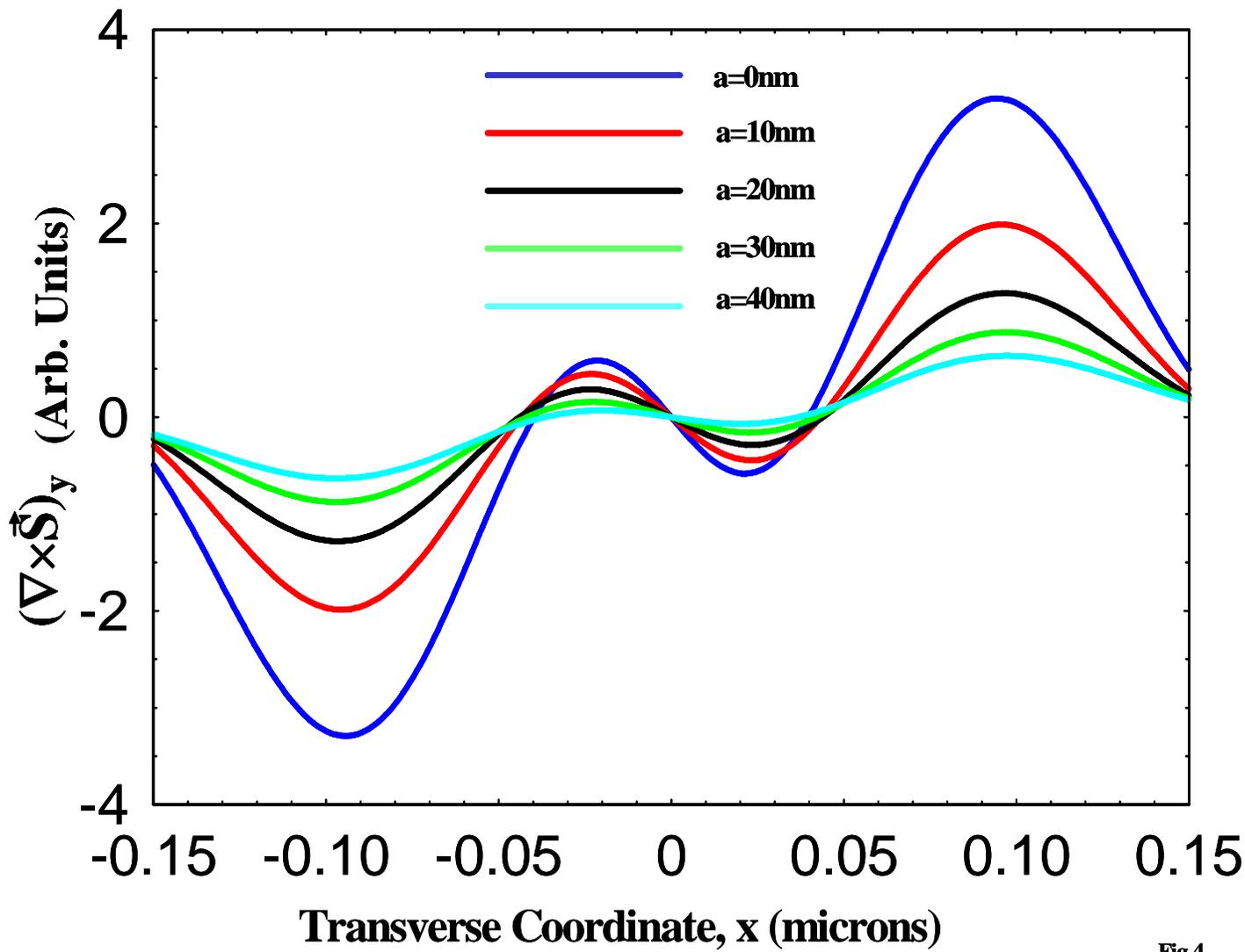

Fig.4

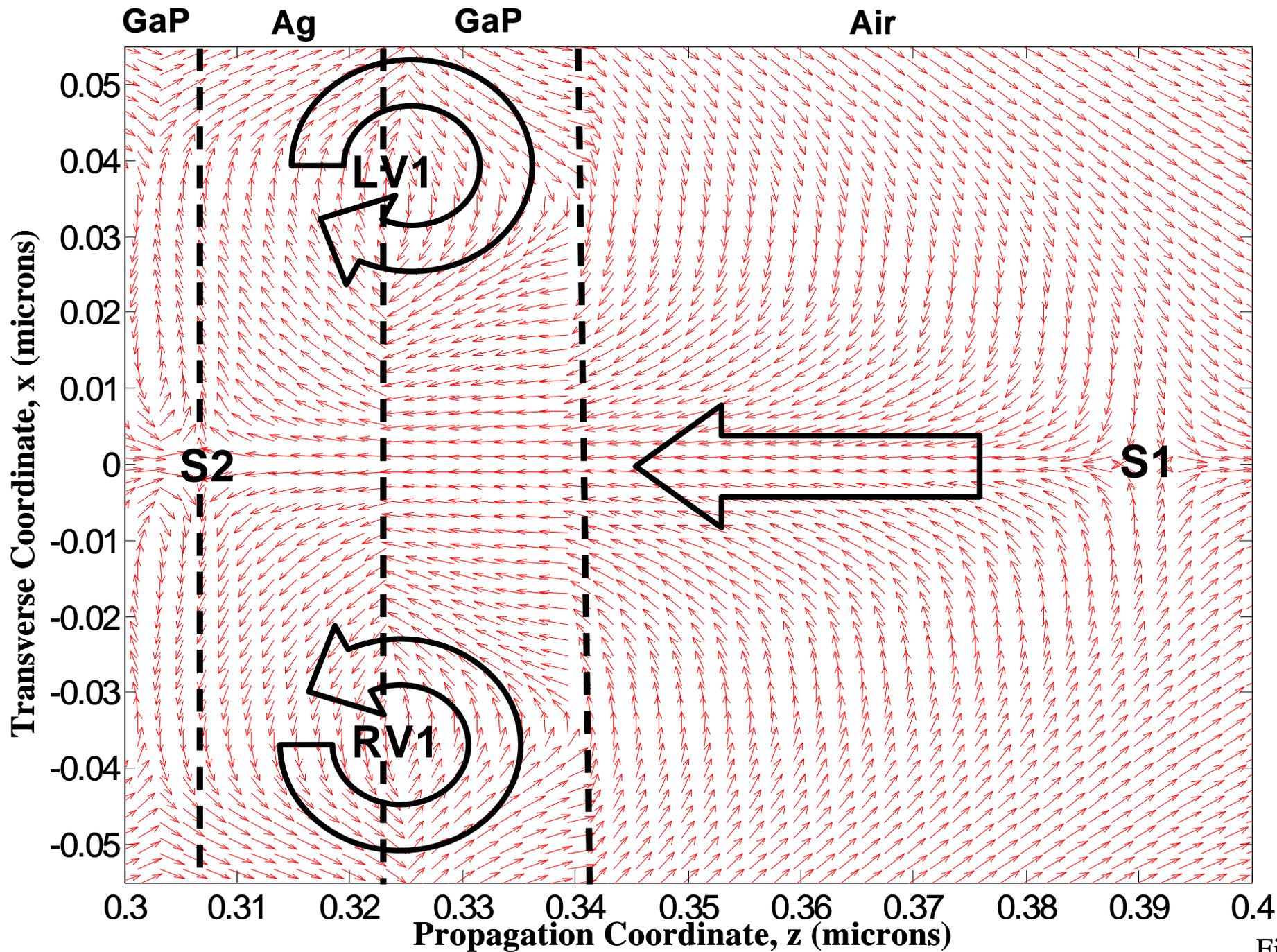

Fig.5